
\input phyzzx
%
%
\def\e{\adveq\eqno{\rm (\chapterlabel\the\equanumber)}}
\def\adveq{\global\advance\equanumber by 1}

%
%
\rightline{IC/92/150}
\date{July 1992}
\titlepage
\vskip 1cm
\title{BRST cohomology ring in 2D gravity coupled to minimal models}
\author {H. Kanno $^a$ \footnote{\dagger}{Address after July 1992:
DAMTP, Univ. of Cambridge.} and M.H. Sarmadi $^{a,b}$}
\address{a) International Centre for Theoretical Physics \break
P.O.Box 586, Trieste, 34100, Italy \break
b) INFN, Sezione di Trieste, Trieste Italy }
\abstract{  The ring structure of Lian-Zuckerman states for $(q,p)$ minimal
models coupled to gravity is shown to be ${\cal R}={\cal R}_0\otimes {\bf C}
[w,w^{-1}]$ where ${\cal R}_0$
is the ring of ghost number zero operators generated by two elements and
$w$ is an operator of ghost number $-1$. Some examples are discussed in detail.
For these models the currents are also discussed and their algebra is shown
to contain the Virasoro algebra.}
%
%
\def\cmp#1{{\it Comm. Math. Phys.} {\bf #1}}

\def\pl#1{{\it Phys. Lett.} {\bf B#1}}
\def\prl#1{{\it Phys. Rev. Lett.} {\bf #1}}

\def\np#1{{\it Nucl. Phys.} {\bf B#1}}

\def\jmath#1{{\it J. Math. Phys.} {\bf #1}}
\def\mpl#1{{\it Mod. Phys. Lett.} {\bf A#1}}
\def\ijmp#1{{\it Int. J. Mod. Phys.} {\bf A#1}}

\REF\BDG{ E. Br\'ezin and V. Kazakov, \pl{236} (1990) 144; \hfill \break
M.R. Douglas and S. Shenker, \np{335} (1990) 635; \hfill \break
D.J. Gross and A.A. Migdal, \prl{64} (1990) 127.}
\REF\DDK{ J. Distler and H. Kawai, \np{321} (1989) 509;  \hfill \break
 F. David \mpl{3} (1989) 1651.}
\REF\SP{ N. Seiberg, {\it Prog.Theor. Phys. Suppl.} {\bf 102} (1991) 319;
 \hfill \break  J. Polchinski, \np{346} (1990) 253, \np{357} (1991) 241.}
\REF\GL{ M. Goulian and M. Li, \prl{66} (1991) 2051;  \hfill \break
P. Di Francesco and D. Kutasov, \pl{261} (1991) 385.}
\REF\KD{ Y. Kitazawa, \pl{265} (1991) 262;  \hfill \break
Vl.S. Dotsenko, \mpl{6} (1991) 3601, preprints, PAR-LPTHE 92-4 (January 1992),
CERN-TH.6502/92,PAR-LPTHE 92-17 (May 1992).}
\REF\ST{N. Sakai and Y. Tanii, {\it Prog. Theor. Phys} {\bf 86}
(1991) 547, \ijmp{6} (1991) 2743, \pl{276} (1992) 41.}
\REF\LZA{B.H. Lian and G.J. Zuckerman, \pl{254} (1991) 417.}
\REF\GK{D.J. Gross and I. Klebanov, \np{359} (1990) 3;  \hfill \break
D.J. Gross, I. Klebanov and M. Newman \np{350} (1990) 333;  \hfill \break
U.H. Danielsson and D.J. Gross, \np{366} (1991) 3.}
\REF\SW{A.M. Sengupta and S.R. Wadia, \ijmp{6} (1991) 1961;  \hfill \break
G. Mandal, A.M. Sengupta and S.R. Wadia, \mpl{6} (1991) 1465.}
\REF\BK{M. Bershadsky and I. Klebanov, \np{360} (1991) 559.}
\REF\PK{ A.M. Polyakov, \mpl{6} (1991) 635;  \hfill \break
I. Klebanov and A.M. Polyakov, \mpl{6} (1991) 3273.}
\REF\LZB{B.H. Lian and G.J. Zuckerman, \pl{266} (1991) 21}
\REF\BMP{P. Bouwknegt, J.M. McCarthy and K. Pilch, \cmp{145}
(1992) 541.}
\REF\MMS{S. Mukherji, S. Mukhi and A. Sen, \pl{266} (1991) 337;  \hfill \break
C. Imbimbo, S. Mahapatra and S. Mukhi, \np{375} (1992) 399.}
\REF\IO{K. Itoh and N. Ohta, preprint FERMILAB-PUB-91/228-T.}
\REF\WIT{E. Witten, \np{373} (1992) 187.}
\REF\KMS{D. Kutasov, E. Martinec and N. Seiberg, \pl{276} (1992) 187.}
\REF\KLE{I.R. Klebanov \mpl{7} (1992) 723.}
\REF\WZ{E. Witten and B. Zwiebach, preprint IASSNS-HEP-92/4
(January 1992).}
\REF\CDK{N. Chair, V.K. Dobrev and H. Kanno, \pl{283} (1992) 194.}
\REF\GJJM{S. Govindarajan, T. Jayaraman, V. John and P. Majumdar,
preprint IMSc-91/40, (December 1991).}

\section{Introduction}

Two dimensional gravity coupled to conformal matter has been
investigated in various approaches as a toy model of quantum gravity.
The matrix model approach has perhaps uncovered some non-perturbative
aspects and in any case gives a powerful computational method combined with
the theory of integrable systems [\BDG].
There is a more conventional method called continuum approach:
in the conformal gauge the induced gravity sector is described by the
Liouville theory [\DDK].
Though the interacting quantum Liouville dynamics itself
is very complicated [\SP], calculations based on a free field realization
have given some results in agreement with the matrix model
approach [\GL,\KD,\ST], in particular for $c=1$.
\par
In the BRST quantization framework, if one takes as matter sector
the minimal model of BPZ, there exists an infinite tower of physical states
(BRST cohomologies) for each conformal block [\LZA].
This result is very remarkable
in the sense that we have a physical state with any ghost number in contrast
with the usual situation. Hence, it is crucial to clarify the origin and
the implications of this infinite tower structure for a better
understanding of 2D quantum gravity coupled to minimal models.
Using the state-operator correspondence,
we can construct physical operators (observables) from physical
states (the Lian -Zuckerman states). The short distance behavior
of the operator product defines a ring structure of BRST
cohomologies. Thus we obtain an associative ring of observables.
\par
For $c=1$ matter, the existence of such discrete
states or extra states is observed in several ways [\GK-\IO].
The significance
of the ring structure, the so-called ground ring, and the
symmetry currents acting on it was pointed out by Witten [\WIT] . It
turns out this algebraic structure is very useful for both
practical calculations of correlation functions and physical
interpretation of the $c=1$ matrix model [\WIT-\WZ].
\par
In this paper, we will discuss the ring structure for minimal
models. An important difference arises from the existence of the
infinite tower of BRST cohomologies
which is absent in the $c=1$  case. We will propose
a ring structure behind this infinite tower.  We show the
existence of a generator $w$ with ghost number $-1$
together with its `inverse'
$w^{-1}$. By proving any power of $w$ gives a non-trivial BRST cohomology,
we find the ghost number $n$ sector is given by
$$
{\cal R}_n~=~w^{-n}~{\cal R}_0~,  \e
$$
where ${\cal R}_0$ is the ring of ghost number zero physical operators.
The ring structure of ${\cal R}_0$ has been discussed in [\KMS] and will be
explained below. Consequently the full ring structure is
$$
{\cal R}~=~{\cal R}_0 \otimes {\bf C} [w,w^{-1}]~,  \e
$$
which is one of main results in the present paper.
As we will see below, this ring is non-commutative; the generator $w$
anti-commutes with the generators of ${\cal R}_0$.
The infinite tower structure is essentially generated by the
single element $w$. We will prove the ring structure (2) for the (2,3), (2,5),
(2,7) and (3,4) minimal models by an explicit calculation of the operator
product expansion.
In general cases we will give an argument based on the
calculation of relevant 3-point functions.
\par
The organization of the paper is as follows.
In Section 2 we briefly review the result of Lian and Zuckerman on physical
states. In Section 3 we discuss how the above ring structure can follow by
considering the Liouville momenta. Section 4 contains some examples for
which the ring structure is proved, and also the algebra of the vector fields
obtained from the currents is discussed. In Section 5 we give a discussion of
the general $(q,p)$ models.
\def\modul{{\cal M}}
\def\del{\partial}
\section{Lian-Zuckerman States and Physical Operators}

In the BRST quantization procedure the physical states are
defined to be BRST cohomology classes. For any conformal
field theory (CFT) with the total central charge $c^{tot}=26$,
we can introduce the (Virasoro) BRST complex with coefficients
in the Virasoro module ${\cal M}^{c=26}$. The BRST operator
$$
Q_B~=~\oint {dz\over 2\pi i}:\big( T(z)+{1\over 2}T^G(z)\big)c(z):, \e
$$
acts on the cochain space
$$
C^*(Vir,\modul)~=~\modul\otimes\Lambda^{bc}~, \e
$$
where $T(z)$ is the energy momentum tensor for $\modul$, and  $\Lambda^{bc}$
is the Fock module of the \hfill \break  reparametrization ghosts $(b,c)$ with
energy momentum tensor $T^G(z)$. The $Z$-gradation
of $C^*(Vir,\modul)$ is called ghost number. Let us consider
the case
$$
\modul~=~L(c_{q,p},\Delta_{m',m}) \otimes {\cal F}^L (Q^L,p^L)~, \e
$$
corresponding to 2D gravity coupled to the $(q,p)$ minimal model.
$L(c_{q,p},\Delta_{m',m})$ is the Virasoro irreducible module
with central charge
$$
c_{q,p}~=~1 - {6(q-p)^2\over qp}~,\qquad (q<p,~coprime) \e
$$
and highest weight
$$
\Delta_{m',m}~=~{1\over 4qp} \bigl[ (qm'-pm)^2 - (q-p)^2 \bigr]~, \e
$$
$$
(1\leq m \leq q-1,~~1\leq m' \leq p-1 )~.
$$
We assume a free field realization of the Liouville field.
${\cal F}^L (Q^L,p^L)$ is the Feigin-Fuchs module with the background
charge $Q^L$ and vacuum momentum $p^L$. The central charge
and the Virasoro highest weight are given by
$$
c~=~1 + 12 (Q^L)^2~, \qquad h(p^L)~=~ -{1\over 2}p^L(p^L - 2Q^L)~. \e
$$
The condition $c^{tot}=26$ determines $Q^L$ up to sign.
For a moment we will focus on the relative BRST cohomology
$H^*_{rel}(Vir,\modul)$, in which we take the subcomplex satisfying
$L_0^{tot}\psi=b_0\psi=0,~~\psi\in C^*(Vir,\modul)$.
\par
If we fix the matter sector, $H^*_{rel}(Vir,\modul)$ depends on
the value of the Liouville momentum $p^L$. Lian and Zuckerman proved that
$H^*_{rel}(Vir,\modul)$ is non-trivial for only special discrete
values of $p^L$ and that there exists a physical state for an arbitrary
ghost number (i.e. an infinite tower of physical states) for each conformal
block.
Let $E_{m',m}(q,p)$ denote the set of highest
weights appearing in the embedding diagram of Verma modules :
$$
E_{m',m}(q,p)~=~\{ a_t,~b_t \}_{t\in Z}~, \e
$$
$$
\eqalign{
a_t &=~ {1\over 4qp} \bigl[ (2qpt + qm' + pm)^2 - (q-p)^2
                         \bigr]~,\cr
b_t &=~ {1\over 4qp} \bigl[ (2qpt + qm' - pm)^2 - (q-p)^2
                         \bigr]~.\cr}\e
$$
The relative cohomology $H^*_{rel}(Vir,\modul)$ is non-vanishing
if and only if the Liouville momentum $p^L$ satisfies
$$
1-h(p^L) \in E_{m',m}(q,p)~. \e
$$
$H^*_{rel}(Vir,\modul)$ is at most one-dimensional. That is we
have a unique physical state (Lian-Zuckerman state)
for each $p^L$ satisfying (11). The ghost number is given by
$$
n_{gh}~=~\pi(p^L)~d(p^L)~, \e
$$
where
$$
\eqalign{
d(p^L) :=&~~~{\hbox{number of arrows leading from}}~~~b_0=\Delta_{m',m}\cr
&~~~~{\hbox{to}}~~~1-h(p^L)~~~{\hbox {in the embedding diagram}}~,\cr} \e
$$
and
$$
\pi(p^L) := sign~(p^L - Q^L)~. \e
$$
Note that for each weight $\Delta \in E_{m',m}(q,p)$ there exist
two momenta such that $1-h(p^L)=\Delta$, which are distinguished from each
other by $\pi(p^L)$. Therefore, preparing two copies of the
embedding diagram, one for the states with $\pi(p^L)>0$ and the other
for $\pi(p^L)<0$, we can make a one-to-one correspondence between
the Lian-Zuckerman states and the nodes of the diagrams .
\par
One can construct physical operators (observables) $\cal O$ from the
Lian-Zuckerman states by the standard prescription in CFT. The translation
to the operators enables us to introduce a ring structure of BRST
cohomology. In the coulomb gas description of the matter sector in terms of
a free field $X(z)$, the matter Virasoro generators are expressed in terms of
the oscillators $\del ^nX$. The Liouville vacuum
is represented by the vertex operator $e^{\beta\phi(z)}~(\beta=p^L)$.
Then the physical operator in general takes the form
$$
{\cal O}~=~{\cal P}\big[ \del X, \del\phi, b,c \big]~\Phi_{m,m'}
          ~e^{\beta\phi}~,   \e
$$
where $\cal P$ is a differential polynomial with a definite conformal
weight and $\Phi_{m,m'}$ is a matter primary field. (In the following
we give explicit examples.)
Compared with the original states, the ghost number of the
operators increases by one due to the gap between the $SL_2$-invariant ghost
vacuum and the highest weight state of $\Lambda^{bc}$. In other words we will
consider the observables in the $0$-form version, not in the $2$-form
version.  Let $H^*_{LZ}~(Vir,(q,p))$ denote the set of observables in
2D gravity coupled to the $(q,p)$ model. Since the $(q,p)$ model consists
of ${1\over 2}(q-1)(p-1)$ primaries or conformal blocks, we have
$$
{\hbox{dim}}~H^n_{LZ}~(Vir,(q,p))~=~(q-1)(p-1)~, \e
$$
for any ghost number $n$. Note that in the above we only discussed the (chiral)
relative cohomology. As we will see below, knowing the states in the
relative cohomology it is easy to obtain the full set of states in the absolute
cohomolgy which has twice as many states.
\par
\section{Ring Structure and the Spectrum of the Liouville Momentum}
\def\chg#1#2{\beta_{#1,#2}}
\def\obs{{\cal O}}
\def\ng{n_{gh}}
The short distance behavior of operator product (OPE) of two observables
$$
\obs_1(z)\obs_2(w)~\sim~\cdots + (z-w)^{-n}A_n(w) + \cdots +
          A_0(w) + O((z-w))~, \e
$$
defines a ring structure of $H^*_{LZ}~=~\oplus_{n\in Z} H^n_{LZ}$. The
expansion coefficient $A_n(w)$ commutes with the BRST charge $Q_B$
and has conformal weight $(-n)$, since $\obs_1$ and $\obs_2$ are BRST
cohomologies with weight $0$. But due to the relation $\big[ Q_B,
b_0 \big]~=~L^{tot}_0$, only $\obs_3\equiv A_0$ may give a BRST
non-trivial operator. Thus we get the ring structure
$$
H^n_{LZ}~\times~H^m_{LZ} \longrightarrow H^{n+m}_{LZ}~,
\quad (n,m\in Z) \e
$$
defined by the OPE modulo BRST exact terms
$$
\obs_1(z)~\obs_2(w)~\sim~\obs_3(w)~+~\big[ Q_B, * \big]~,  \e
$$
which will be denoted simply by
$$
\obs_1~\cdot~\obs_2~=~\obs_3~. \e
$$
The fundamental OPE of the vertex operators of the free scalar
field:
$$
:e^{\alpha\phi(z)}:~:e^{\beta\phi(w)}:~=~(z-w)^{-\alpha\beta}
                  :e^{\alpha\phi(z)+\beta\phi(w)}:~, \e
$$
is valid by assumption of a free field realization of the Liouville field.
The OPE (21) implies the conservation of the Liouville charge under the
ring multiplication.
The product of two cohomologies $\obs_1 [\beta_1],\obs_2 [\beta_2]$ with
charges $\beta_1$ and $\beta_2$ can generate another cohomology
$\obs_3$ with charge $\beta_1 +\beta_2$:
$$
\obs_1 [\beta_1]\cdot \obs_2 [\beta_2]~=~\obs_3 [\beta_1 +\beta_2]~. \e
$$
Hence, the whole `spectrum' of the Liouville charge of observables in
$H^*_{LZ}$ gives some insight into a possible choice of generators
of the cohomology ring.
\par
Let us introduce the standard notation for the Liouville charges;
$$
\beta_+:=~\sqrt{{2p\over q}}~, \quad \beta_-:=~\sqrt{{2q\over p}}~,
\quad (\beta_+\beta_-~=~2) \e
$$
$$
\eqalign{
\beta_0:=&~\beta_+~+~\beta_-~=~2Q^L~,  \cr
\chg{n'}{n}:=&~{1\over 2} \bigl[ (1-n)\beta_+~+~(1-n')\beta_- \bigr]~, \cr}  \e
$$
$\chg{n'}{n}$ satisfies the following relation,
$$
\eqalign{
\chg{n'\pm p}{n\mp q}~=&~\chg{n'}{n}~, \cr
\chg{n'}{n} + \chg{k'}{k}~=&~\chg{n'+k'-1}{n+k-1}~.  \cr}\e
$$
Solving the equation
$$
\eqalign{
1 +& {1\over 2}\beta(\Delta) \big( \beta(\Delta) - \beta_0 \big)~
            =~\Delta~, \cr
\Delta~=&~a_t,~b_t~\in E_{m',m}(q,p)~, \cr}\e
$$
we obtain the following `spectrum' of the Liouville charge:
$$
\eqalign{
\beta^\pm (a_t)~=&~{1\over 2} \bigl\{ \beta_0 \pm \vert
    (tq+m)\beta_+ + (tp+m')\beta_- \vert \bigr\}~, \cr
\beta^\pm (b_t)~=&~{1\over 2} \bigl\{ \beta_0 \pm \vert
    (tq-m)\beta_+ + (tp+m')\beta_- \vert \bigr\}~, \cr}\e
$$
where the two branches of the solution are given by the sign in
front of the absolute value. Let $k$ be a positive integer. The
ghost number $\ng$ for each Liouville charge in (27) is given
by (cf. Eq.(12)):
$$
\eqalign{
\beta^+(a_{k-1}),~~~\beta^+(a_{-k}) &\longrightarrow \ng = 2k~,\cr
\beta^+(b_{k}),~~~\beta^+(b_{-k}) &\longrightarrow \ng = 2k+1~,\cr
\beta^-(a_{k-1}),~~~\beta^-(a_{-k}) &\longrightarrow \ng = -2k+2~,\cr
\beta^-(b_{k}),~~~\beta^-(b_{-k}) &\longrightarrow \ng = -2k+1~,\cr
\beta^+(b_0),~~~\beta^-(b_0) &\longrightarrow \ng = 1~.\cr	}\e
$$
Removing the absolute value by the condition $1\leq m \leq q-1,~
1\leq m' \leq p-1$, we see the Liouville charges are
$\chg{m'-np}{m-nq}$ and $\chg{(p-m')-np}{(q-m)-nq}$ for even ghost
number $(\ng =2n)$. On the other hand, for odd ghost number
$(\ng =2n+1)$, the charges are $\chg{m'-np}{-m-nq}$
and $\chg{-m'-np}{m-nq}$. Note that the $\ng =1$ case reproduces
the usual gravitational dressing $\chg{m'}{-m}$
and $\chg{-m'}{m}$ for the primary field.
Making use of the relation (25) for $\chg{n'}{n}$, we can
parametrize the $(q-1)(p-1)$ Liouville charges of observables with
ghost number $\ng$ as follows:
$$
\eqalign{
\beta(s,s')~=&~-\ng~\beta_w~+~s'\beta_x~+~s\beta_y~, \cr
  (0\leq s &\leq q-2~, \quad 0 \leq s' \leq p-2)~, \cr}\e
$$
where we have defined
$$
\eqalign{
\beta_w~:=&~\chg{p+1}{1}~=~\chg{1}{q+1}~, \cr
\beta_x~:=&~\chg{2}{1}~, \quad \beta_y~:=~\chg{1}{2}~. \cr}\e
$$
We will restrict ourselves to matter blocks whose labels are inside
the Kac-table. Note that the labels of some Liouville charges, for instance
$\beta _w$, can sit outside the table.
\par
Let $w,x$ and $y$ denote the observables with the charges $\beta_w,
\beta_x$ and $\beta_y$, respectively. $x,y \in H^0_{LZ}$ and
$w \in H^{-1}_{LZ}$. Then the linear spectrum (29) means a possible
identification
$$
\obs~\big[ \beta(s,s') \big]~=~w^{-\ng}x^{s'}y^{s}~, \e
$$
for the observables with ghost number $\ng$, {\it if the right hand side
does not vanish modulo BRST exact terms.}
The generators $x$ and $y$ belonging to the ghost number zero sector
can be obtained by an $SO(2,C)$ rotation from the generators of the chiral
ground ring for $c=1$ matter [\KMS,\CDK]. In the Coulomb gas description
of the matter sector, they are given by the following expressions:
$$
\eqalign{
x&=\bigl( bc+(i\alpha_{1,2}\del X+\beta_{1,2}\del \phi)\bigr)
       e^{i\alpha_{2,1} X+\beta_{2,1} \phi}~,\cr
y&=\bigl( bc+(i\alpha_{2,1}\del X+\beta_{2,1}\del \phi)\bigr)
       e^{i\alpha_{1,2} X+\beta_{1,2} \phi}~,\cr}
$$
where $\alpha_{n',n}={{1-n}\over 2}\alpha_+~+~{{1-n'}\over 2}\alpha_-,~
\alpha_+=\beta_+$ and $\alpha_-=-\beta_-$. (The background matter charge
is equal to $-\alpha_0$ where $ \alpha_0=\alpha_+~+~\alpha_-$.) It is easy to
see that $x^{p-1}$ and $y^{q-1}$ have matter charges which are not
inside the Kac-table and if one is restricting oneself to only those
operators whose charges $\alpha_{m',m}$ are inside the table, i.e.
$1\leq m\leq q-1$ and $1\leq m'\leq p-1$, then one should set
$x^{p-1}=y^{q-1}=0$. Therefore ${\cal R}_0$ has a simple structure with some
connection with the chiral ring for $c=1$ matter.
The novel feature of the minimal models
arises from the generator $w^{\pm 1}$ with $\ng=\mp 1$. To establish the
identification (31) we have to prove that the product of generators
indeed give BRST non-trivial cohomologies, which is the problem we will be
concerned with in the following sections.
\section{Examples}
In this section we will discuss the above ring structure in some examples.
\par
\noindent {\it {(2,3) Model = Pure Liouville Gravity}}\hfill \break
Let us take the $(2,3)$ model ($c_{2,3}=0$) as a simple example. We will
see by explicit calculations that the ring structure of $H^*_{LZ}(Vir,(2,3))$
is exactly the one proposed in the last section. In this case the only
matter primary is the identity $(1,1)=(1,2)$ sector. The matter Verma
module $M(c=0,\Delta =0)$ has its first two singular states at level one
and two $(a_0=1,a_{-1}=2)$.
Hence, we can take a representative of BRST cohomologies in the
irreducible module such that there is no matter oscillator. (Recall that
$L_{-n}~(n>0)$ is generated by $L_{-1}$ and $L_{-2}$.) In this sense the
matter part is trivial as expected and we can think
of it as pure Liouville gravity. The basic parameters for the Liouville
charges are
$$
\eqalign{
\beta_+~&=~\chg{1}{-1}~=~\sqrt 3~, \quad
\beta_-~=~\chg{-1}{1}~=~{2\over\sqrt 3}~, \cr
\beta_x~&=~\chg{2}{1}~=~-{1\over\sqrt 3}~, \quad
\beta_w~=~\chg{1}{3}~=~-\sqrt 3~, \cr}\e
$$
\par
The Lian-Zuckerman states with $\ng=-1$ exist at level one and two.
The one with charge $\chg{1}{1}=0$ gives the identity operator $\bf 1$.
By an explicit construction of the other state with charge $\chg{2}{1}$,
we get a ghost number zero observable
$$
x~=~\big( bc~-~{\sqrt 3\over2}\del\phi \big)~e^{-{1\over\sqrt3}\phi}~.\e
$$
It is the existence of a vanishing null vector at level two
$$
\big( L_{-2}~+~{3\over2}{L_{-1}}^2\big)~
\vert p^L=-{1\over\sqrt 3}\rangle~, \e
$$
which makes $x$ be in the kernel of $Q_B$.
The observable $w$ with charge $\chg{1}{3}$
comes from the Lian-Zuckerman state with $\ng=-2$
at level $b_1=5$. Explicitly,
$$
w~=~\big( b\del b c~-~{1\over\sqrt3} b\del^2 \phi
       ~+~{1\over 2\sqrt3} \del b\del\phi ~+~{1\over6}\del^2 b
        \big)~e^{-\sqrt3\phi}~. \e
$$
The vanishing null vectors at level 3 and 4,
$$
\eqalign{
\big( L_{-3}~+~L_{-1}L_{-2}~+~{1\over6}{L_{-1}}^3\big)~
&\vert p^L=-{\sqrt 3}\rangle~, \cr
\big( L_{-4}~+~{L_{-2}}^2~+~{2\over3}{L_{-1}}^2L_{-2}
{}~+~{1\over12}{L_{-1}}^4\big)~
&\vert p^L=-{\sqrt 3}\rangle~, \cr} \e
$$
are responsible for the BRST invariance of $w$.
On the other hand, the observables with $\ng=1$ are nothing but the
Liouville `screening' operators multiplied by the $c$-ghost:
$$
s_+~=~ce^{\beta_+\phi}~, \quad s_-~=~ce^{\beta_-\phi}~. \e
$$
By computing the OPE, we see
$$
s_+(z)w(0)~\sim~-{1\over6}\cdot{\bf 1}~, \quad
s_-(z)w(0)~\sim~-{1\over3}x(0)~. \e
$$
This proves the identification $s_+=-{1\over 6}w^{-1}$ and
$s_-=-{1\over 3}xw^{-1}$. For each fixed ghost number
we have exactly two observables in the relative cohomology.
The very existence of the inverse
$w^{-1}$ in the cohomology ring $H^*_{LZ}$ is enough to prove that
for any $n\in Z$, neither $w^n$ nor $w^n x$ vanishes modulo BRST exact terms.
Therefore, $x\in H^0_{LZ}$ and $w\in H^{-1}_{LZ}$ are two of the generators of
the ring. For example, we can obtain the other ghost number $-1$
observable by taking the OPE coefficient:
$$
\eqalign{
w(z)x(0)~\sim~&\biggl[ {1\over48} \del^4 b~-~{2\over9}\del^3b(bc)
            ~+~{3\over4}\del^2 b(\del bc)~+~{1\over4}\del b(b\del^2 c)
            ~+~{1\over2\sqrt 3}\del^2 b(bc)\del\phi \cr
			&-{\sqrt 3\over4}\del^2 b\del^2\phi~+~\del b(bc)
			 \big( {5\over2\sqrt 3}\del^2\phi~-~{2\over3}(\del\phi)^2 \big)
	 ~+~{1\over8\sqrt 3}\del b\big(\del^3\phi~+~2(\del\phi)^3\big) \cr
	 &+b\bigl({-1\over12\sqrt 3}\del^4\phi~+~{1\over2}(\del^2\phi)^2~+~
	   {1\over12}\del^3\phi\del\phi~-~{1\over2\sqrt 3}\del^2\phi
	   (\del\phi)^2 \bigr) \biggr]
	     \exp\bigl( {-4\over\sqrt 3} \phi\bigr) ~, \cr} \e
$$
which corresponds to the Lian-Zuckerman states at level $b_{-1}=7$.
\par
So far we considered only the relative cohomology. It contained the
operators $w^n$ and $xw^n$ ($n\in Z$). The absolute cohomology contains twice
as many physical operators. As in the $c=1$ case of ref. [\WZ],
the other half of the operators are conveniently obtained by multiplying
the above operators by the physical operator
$$
a~=~c\del \phi+{1\over 2} \beta _0\del c ~. \e
$$
Therefore the full set of chiral operators are of the form  $w^n,xw^n,aw^n$ and
$ axw^n$. Note that $aw=-wa$ which is due to  $a$ and $w$
both carrying nonzero ghost numbers, and $xw=-wx$ which, even though
$x$ has ghost number zero, is due to the anticommutation of the exponentials.
However $ax=xa$. By introducing appropriate cocycle factors, say the Pauli
matrices $\sigma ^i$, one can instead have two anti-commuting generators,
${\tilde x},{\tilde a}$ and one commuting one ${\tilde w}$. As we will see
shortly it is more natural to use these latter generators.
\par
Using the above operators we can construct currents and the corresponding
charges which act on the cohomology ring as derivations. If $\psi ^{(n)}$ is
a physical operator of ghost number $n$ then $j^{(n-1)}(z)=\oint dz'
b(z')\psi (z)$ is an operator of ghost number $(n-1)$ and conformal
weight one whose BRST variation is a total derivative. Therefore for some other
physical operator ${\tilde \psi}$ the BRST variation of of $\oint
dzj^{(n)}{\tilde \psi}^{(m)}$ vanishes and if it is not exact then it
corresponds to
a physical operator of ghost number $(m+n)$. For example, the operator
$s_-=ce^{\beta_-\phi}$ gives rise to the current $e^{\beta_-\phi}$ and
by considering the action of $\oint dze^{\beta_-\phi}$ on physical
operators one sees that it should be identified with the vector field
$xw^{-1}\partial _a$. Similarly from the operator $w$ one obtains the
current $b\partial be^{-\sqrt{3} \phi}$ whose integral should be identified
with the vector $w\partial _a$. By considering the other physical
operators one can show that one has  the following four sets of vector
fields:
$$
\eqalign{
&G_n=w^n\partial _a ,\cr
&H_n=xw^n\partial _a , \cr
&L_n=-w^n(w\partial _w+1/3 x\partial _x-na\partial_a), \cr
&K_n=-xw^n(w\partial _w-(n+1/3)a\partial _a).\cr} \e
$$
The above vector fields are obtained respectively from the operators
$w^n, xw^n, aw^n$ and $axw^n$. The generators $L_n$ satisfy a
Virasoro algebra $[L_n,L_m]=(n-m)L_{n+m}$. In order to write the other
commutators in a more recognizable form, we first write the vector fields
in terms of ${\tilde w},~{\tilde x}$ and ${\tilde a}$ which we denote by
${\tilde L},{\tilde K},{\tilde G}$ and
${\tilde H}$. These set of generators satisfy the following algebra:
$$
\eqalign{
[{\tilde L}_n,{\tilde L}_m]&=(n-m){\tilde L}_{n+m},\cr
[{\tilde L}_n,{\tilde K}_m]&=(n-m+1/3){\tilde K}_{n+m},\cr
[{\tilde L}_n,{\tilde G}_m]&=-(n+m){\tilde G}_{n+m},\cr
[{\tilde L}_n,{\tilde H}_m]&=(-n-m-1/3){\tilde H}_{n+m},\cr
[{\tilde K}_n,{\tilde K}_m]&=0,\cr
[{\tilde K}_n,{\tilde G}_m]&=-(n+m+1/3){\tilde H}_{n+m},\cr
[{\tilde K}_n,{\tilde H}_m]&=0. \cr } \e
$$
Therefore, the vectors ${\tilde K},{\tilde G}$ and ${\tilde H}$
are primaries of the Virasoro
generators ${\tilde L}_n$ with weights 2,0 and 0 respectively. Moreover
${\tilde G}$ has integer modes but ${\tilde K}$ and ${\tilde H}$ have $n+1/3$
and $n-1/3$
modings. The vanishing of the commutator of ${\tilde K}$ with itself and with
${\tilde H}$ is simply because $x^2=0$.

\noindent {\it (2,5) Model }\hfill \break
\def\dpol#1{{\cal L}_{-#1}}
Our next example is the $(2,5)$ model $(c_{2,5}=-{22\over5})$,
the Lee-Yang edge singularity model coupled to gravity. This is
complicated enough, since the matter field enters in non-trivial way.
Our strategy to prove the ring structure is the same as before.
The generator $w$ is constructed from the Lian-Zuckerman state at
level 7 in the $(1,1)=(1,4)$ sector. In the course of finding the
Lian-Zuckerman state we must use the existence of vanishing null
vectors at levels 3 and 6 in the Liouville Fock module. We
also use the fact that there are singular vectors at level 1 and
4 in the identity sector of the matter Verma module. The BRST
variation of the Lian-Zuckerman state is a linear combination of
descendents of these singular vectors. (Of course, the
Lian-Zuckerman state is defined up to descendents of singular
vectors in addition to the freedom of BRST exact terms.)
Expanded in the ghosts, $w$ takes the form
$$
\eqalign{
w~=~&\biggl[ {a_1\over4!}\del^4 b~+~{a_2\over3!}\del^3 b(bc)~+~
       {a_3\over2!}\del^2 b(\del bc)~+~{1\over 3!}\dpol1\del^3b
	   ~+~{1\over 2!}\widetilde{\dpol1}\del^2b(bc) \cr
	   &~+~{1\over 2!}\dpol2\del^2b~+~\widetilde{\dpol2}\del b(bc)
  ~+~\dpol3\del b ~+~\dpol4 b \biggr]~\exp(-\sqrt 5\phi)~,\cr} \e
$$
where $5a_1~+~3a_2~+~a_3~=0$ and $\dpol{n}~(\widetilde{\dpol{n}})$
is a differential polynomial in both the matter and the
Liouville fields with conformal weight $n$.
The explicit form is presented in the appendix.
$\dpol{n}$ depends on the parameters $a_1$ and $a_2$, one of which
corresponds to an overall factor. We have a one-parameter family
of BRST cohomologies. The point $3a_1+2a_2=0$ gives a BRST trivial case.
(to be precise, we have already fixed one freedom of adding an exact
term by requiring that there is no $\del c(b\del b)$-term.)
Hence the BRST cohomology class is unique in accord with the
general theorem of Lian and Zuckerman.
\par
Among the gravitationally dressed primaries in $H^{+1}_{LZ}$, we take
$$
\obs_{\bf 1}~=~c e^{{\sqrt 5}\phi}~, \quad
\obs_{\sigma}~=~c e^{{4\over\sqrt 5}\phi +{i\over\sqrt 5}X}~,\e
$$
which are in the identity and the spin operator sectors, respectively.
A little tedious calculation of OPE's gives the following results:
$$
\eqalign{
\obs_{\bf 1}\cdot w~=&~-{3\over16}(3a_1 + 2a_2) \cdot {\bf 1}~,\cr
\obs_{\sigma}\cdot w~=&~-{3\over16}(3a_1 + 2a_2) \cdot
          \big( bc - {\sqrt 5\over2}(\del\phi + i\del X)\big)
		   e^{{1\over\sqrt 5}(iX -\phi)}~.\cr} \e
$$
The OPE coefficients vanish if and only if $w$ is BRST trivial. For BRST
non-trivial $w$, the OPE (45) again proves the existence of
$w^{-1} = \obs_{\bf 1}$ in the cohomology ring.
Futhermore, in the second OPE coefficient we recognize the generator
$x \in H^0_{LZ}$.

In order to write the operator $w$ or to check the non vanishing of the
product $\obs_{\bf 1}\cdot w~$, we could have alternatively tried to write
down the vector feild
$\oint dzj^{(-2)}=w\partial _a$ and then consider its action on $a$ or
$aw^{-1}$. In fact this is computationally simpler, and for the next two models
for which the expressions become more involved we will do this. The current
$j^{(-2)}$ can be obtained by demanding that its BRST variation be a total
derivative up to null states. For the present model it has the expression:
$$
j^{(-2)}(z)=\biggl[\bigl[({{-3}\over{2\sqrt{5}}}\del ^2\phi-
           {1\over 2}(\del\phi)^2)-2({{3i}\over{2\sqrt{5}}}\del ^2X-{1\over2}
           (\del X)^2)\bigr]b\partial b
          +{7\over 30} b\partial ^3b -{3\over 10}\partial b \partial ^2b
          \biggr]e^{-\sqrt{5}\phi}. \e
$$
Note that $j$ does not contain the freedom in the choice of the
cohomology class. This is due to the fact that in this case for the exact
states $Q\chi$ where $\chi=L_{-1}b_{-1}b_{-2}b_{-3}c_1|-\sqrt{5}>$ or
$\chi=b_{-1}b_{-2}b_{-4}c_1|-\sqrt{5}>$, we have $b_{-1}Q\chi=L^{total}_{-1}
\chi$ because $b_{-1}\chi=0$.

\noindent {\it (2,7) model}\hfill \break
In this case, $c_{2,7}=-{68\over 7}$, the generator $w$ is
constructed from the states at
level 9 in the (1,1)=(1,6) sector. Since the expression for $w$ contains
many terms, it is more convenient to write the current for
$w\partial _a$ and then show that its action on
$a(z)c(0)e^{\sqrt{7}\phi}=-{{5}\over {2\sqrt{7}}}\partial c ce^{\sqrt{7}\phi }$
is nonzero.
As we mentioned above, this current is obtained by requiring that its
BRST variation to be a total derivative up to matter and Liouville null
states. These null states are at level 6 for the matter sector and level 3
for the Liouville sector. For the current one obtains the following
expression:
$$
\eqalign{
j^{-2}(z)=\bigl[&G_{-4}b\del b+{1\over{2!}}G_{-3}b\del ^2b+{1\over{3!}}
          G_{-2}b\del ^3b+{1\over{5!}}(-{17\over7}\eta+{297\over 49}\epsilon)
          b\del ^5b  \cr
          &+{1\over {2!}}{\tilde G}_{-2}\del b\del ^2b+{1\over{4!}}
          ({8\over 7}\eta-{90\over 49}\epsilon)\del b\del ^4b
            +{1\over{2!3!}}(-{9\over 7}\eta +{105\over 49}\epsilon)
             \del ^2b\del ^3b\bigr]e^{-\sqrt{7}\phi} \cr} \e
$$
The explicit expressions for $G_{-n}$ and ${\tilde G}_{-n}$ are given in
the appendix. In this expression, setting $\epsilon =0$ one obtains a
BRST-trivial current.
The action of $\oint j^{(-2)}dz $ on $aw^{-1}$ gives $-{18\over 49}\epsilon$
which again verifies that $w^{-1}=-{{49\sqrt{7}}\over {45\epsilon}}
ce^{\sqrt{7}\phi }$.
\par
All the above examples were models of type $(2,p)$. For these models the
ring is generated by $w,x$ and $a$ where $x^{p-1}=0,~wa=-aw$ and $wx=-xw$.
The vector fields in eq.(41) which we obtained for the (2,3) model are easily
generalized to $(2,p)$ models as follows. The ones obtained from the operators
$x^iw^n$ have the form
$$
K^{(i)}_n=-x^iw^n\big(w\del _w+{1\over p}x\del _x-
           (n+{i\over p})a\del _a\big)~,\e
$$
and the ones obtained from the operators $ax^iw^n$ are
$$
G^{(i)}_n=x^iw^n\del _a~. \e
$$
After introducing the appropriate cocycle factors as in the (2,3) model, we
find the following algebra:
$$
\eqalign{
[{\tilde K}^{(i)}_n,{\tilde K}^{(j)}_m]&=(n-m+{{i-j}\over p}){\tilde
K}^{(i+j)}_{n+m}~, \cr
[{\tilde K}^{(i)}_n,{\tilde G}^{(j)}_m]&=-(n+m+{{i+j}\over p}){\tilde
G}^{(i+j)}_{n+m}~, \cr
[{\tilde G}^{(i)}_n,{\tilde G}^{(j)}_m]&=0 ~ . \cr }
$$
Therefore ${\tilde K}^{(0)}_n$ generate a Virasoro algebra under which
${\tilde K}^{(i)}_n$ and ${\tilde G}^{(i)}_m$ are primaries of weights two
and zero respectively.
\par
\noindent {\it (3,4) model}\hfill \break
As a final example we consider the Ising model coupled to gravity. Again
we will write the current $w\partial _a$ where in this case $w$ is made out of
states at level 7 and it is in the energy sector (2,1)=(1,3). The null states
are at levels 2 and 3 for the matter sector and at levels 4 and 5 for the
Liouville sector. We obtain the following expression for the current:
$$
\eqalign{
j^{(-2)}(z)=&\biggl[\bigl[(-{5\over 2\sqrt{6} } \partial^2\phi
              -{1\over 2}(\partial\phi)^2)
              -{11\over 9}(-{3\over 2\sqrt{6}}~i\partial^2X
                  -{1\over 2}(\partial X)^2) \bigr]b\partial b \cr
              &-{4\over 3\sqrt{6}}~i\partial X b\partial^2 b
            +{1\over 3} b\partial^3 b-{11\over 12}\partial b\partial^2b\biggr]
             e^{-2/\sqrt{6}~iX-\sqrt{6}\phi}. \cr} \e
$$
The action of $\oint j^{(-2)}dz $ on the state $a(z)c(0)e^{3i/\sqrt{6}X
+\sqrt{6}\phi}=-{5\over 2\sqrt{6}}~\partial cc
e^{3i/\sqrt{6}X+\sqrt{6}\phi}$ gives ${1\over 2}e^{i\alpha_0 X}$,
which again verifies that
$w^{-1}=-4\sqrt{6}/5~ce^{3i/\sqrt{6}X+\sqrt{6}\phi}$. Therefore,
for this model the states are of the type $x^iy^jw^n$ and $ ax^iy^jw^n$
$(n\in Z)$ where $x^3=0$ and $y^2=0$.

\section{Discussion}
For the proof of the ring structure
$$
{\cal R}~=~{\cal R}_0 \otimes {\bf C} [w,w^{-1}]~,  \e
$$
the existence of the `inverse' $w^{-1}$ is a crucial point.
The spectrum of the Liouville momentum discussed in Sect.3 tells us that
there is an observable $v$ in $H_{LZ}^{+1}$
with the desired momentum $-\beta_w$.
The product $w\cdot v$ has vanishing momentum and,
hence, is proportional to the identity. The problem is whether the
proportionality constant is non-vanishing or not. In the general
$(q,p)$ model, by looking at the charge $\beta_w$, we see
that the observable $w$ belongs to the $(1,p-1)=(q-1,1)$ sector
of the matter conformal block. (Note that $w$ is not necessarily in
the identity sector.) However, an explicit construction of the
observable $w$ gets involved in general, since we have to manage
singular vectors at higher levels. We cannot make an OPE
calculation without knowing the form of $w$ explicitly.
Instead of OPE coefficients, however, one may consider an appropriate
three point function on the sphere. If there exists a non-vanishing
three point function containing $w$ and $v$, the proportionality
constant cannot vanish and we can identify $v$ with the inverse
of $w$. By ghost number counting the three point function
$\langle w v \obs \rangle$ can be non-zero only for observables
$\obs$ with ghost number three. For such a correlation function
with matching ghost number, we can reduce it to
a correlation function of Dotsenko-Kitazawa type by making use of
the descent equation trick discussed in [\GJJM].
 This descent equation comes from the double complex consisting
of two coboundary operators, the BRST operator $Q_B$ and the operator
$Q_F$ of Felder's resolution of the Virasoro irreducible module. Noting
that the Liouville charge does not change in `descending' the descent
equation, we can identify relevant Dotsenko-Kitazawa type
operators which are products of matter and Liouville vertex operators.
(The matter momentum is not restricted to the inside of the Kac table.)
If one uses only the operators in relatve cohomology then after using
the descent equations only the operators of Dotsenko-Kitazawa type will
appear and then one can use the results of ref. [\KD] on the calculation
of these types of correlators to conclude that $\langle w v \obs \rangle$
is non-vanishing.
\par
In ref. [\KD], in calculating the correlators, a continuation to a negative
number of screening operators had to be performed.  Since this issue is not
well
understood, it is preferable to modify the above arguement such that
no insertions of screening operators are required.
In ref. [\GJJM], in relating the correlators of Lian-Zuckerman operators
to those of Dotsenko-Kitazawa operators, only the states in relative
cohomology were used. However, if one does not restrict oneself to
using the states which are only in relative cohomology then it is possible
to write correlators which require no screening operators. Let us consider
the general $(q,p)$ minimal model. The operator
$$
{\cal O}^{(3)}=av^2x^{p-2}y^{q-2} ,
$$
which has ghost number three and matter and Liouville charges equal to
$\alpha_0$ and $\beta_0$, has a non-vanishing one-point function.
In order to show that the three point function
$\langle w~v~{\cal O}^{(3)}\rangle$
is non-zero, we first use the descent equation
$$
\del ^2cce^{\beta_0\phi}=[Q_B~,~c\del Xe^{\beta_0\phi}],
$$
then the action of $Q_B$ on $w$ gives:
$$
[Q_B~,~w]=[Q_F~,~\Phi^{(0)}],
$$
where $\Phi^{(0)}$ is an operator of ghost number zero and whose matter charge
is outside the Kac-table. In the above equation one has two choices for $Q_F$,
namely $Q_+$
and$Q_-$ where $Q_{\pm}=\oint e^{i\alpha_{\pm}X}$ are the two screening
operators in the matter sector. Choosing $Q_F=Q_-$ then, by a proper choice of
the representative for $w$, one can take $\Phi^{(0)}=x^p$
\footnote{\ddag}{In the
previous sections, when defining the ring we set $x^{p-1}=y^{q-1}=0$ since
in the Coulomb gas description of the matter sector $x^{p-1}$ and $y^{q-1}$
have charges outside the Kac-table and we were restricting ourselves only
to operators inside the table. However, when discussing Felder's resolution
one needs to allow for operators like $x^p$ and $y^q$ whose charges are outside
the table.} ( The choice $Q_F=Q_+$ corresponds to taking $\Phi^{(0)}=y^q$.)
Now the action of $Q_F$ on $c\del Xe^{\beta_0\phi}$ gives the state
$ce^{i\alpha_{-1,1}X+\beta_0\phi}$. Moreover, it is easy to show:
$$
x^p~v~\sim ~ce^{i\alpha_{1,-1}X}.
$$
Therefore, we have reduced the above three-point function to the two-point
function
$$
\langle \alpha_{1,-1},0|~c_{-1}c_0c_1~|\alpha_{-1,1},\beta_0 \rangle ,
$$
which is obviously non-zero and thus implies that the product $v.w$ does not
vanish.
\par
Finally we note that for the general $(q,p)$ model, the vector fields
${\tilde K}^{i,j}_n$ and ${\tilde G}^{i,j}_n$ which are obtained from
the operators $ax^iy^jw^n$ and $x^iy^jw^n$ should have the froms \footnote
{*}{Recall from the previous section that ${\tilde K}$ and ${\tilde G}$
have the same expressions as $K$ and $G$ but written in terms of
${\tilde w},~{\tilde x},~{\tilde y}$ and ${\tilde a}$.
These latter generators include the cocycle
factors. The original generators of the ring satisfy the (anti-)commutations
$wx=-xw,~wy=-yw,~wa=-aw$ and $xy=yx$. After including the cocycle factors,
${\tilde w}$ is a commuting generator whereas ${\tilde x},~{\tilde y}$ and
${\tilde a}$ are anti-commuting ones.}:
$$
K^{i,j}_n=-x^iy^jw^n\big(w\del _w+{1\over p}x\del _x+{1\over q}y\del _y
          -(n+{i\over p}+{j\over q})a\del _a\big)~,\e
$$
and
$$
G^{i,j}_n=x^iy^jw^n\del _a~. \e
$$
In addition to the $(2,p)$ models discussed in the previous section, we have
checked these expressions for a few currents of the other models. These
vector fields ${\tilde K}^{0,0}_n$ satisfy a Virasoro algebra.

We thank V.K. Dobrev, E. Gava, S. Govindarajan, K.S. Narain and C. Vafa
for discussions. We thank also Professor Salam for hospitality at ICTP.
\vfill \eject

\centerline{\bf Appendix}

The generator of ghost number -1 in the (2,5) model has the form:
$$
\eqalign{
w~=~&\biggl[ {a_1\over4!}\del^4 b~+~{a_2\over3!}\del^3 b(bc)~+~
       {a_3\over2!}\del^2 b(\del bc)~+~{1\over 3!}\dpol1\del^3b
	   ~+~{1\over 2!}\widetilde{\dpol1}\del^2b(bc) \cr
	   &~+~{1\over 2!}\dpol2\del^2b~+~\widetilde{\dpol2}\del b(bc)
         ~+~\dpol3\del b ~+~\dpol4 b \biggr]~\exp(-\sqrt 5\phi)~.\cr}
$$
where
$$
5a_1~+~3a_2~+~a_3~=0~.
$$
Here we the expressions for $\dpol{n}$:
$$
\eqalign{
\dpol1 =&~-{\sqrt 5\over12}(7a_1 + 3a_2)\del\phi~,\cr
\widetilde{\dpol1} =&~{\sqrt 5\over6}(10a_1 + 3a_2)\del\phi~, \cr
\dpol2 =&~-{1\over16}(25a_1 + 22a_2)\big(
           -{3\over2\sqrt 5}\del^2\phi -{1\over2}(\del\phi)^2  \big)
	~+~{1\over8}(25a_1 + 14a_2)\big(
           {3\over2\sqrt 5}i\del^2 X -{1\over2}(\del X)^2
               \big) \cr
		&~-{3\over32}(3a_1 + 2a_2)\big( -{\sqrt 5}\del^2\phi~+~
		    5(\del\phi)^2 \big)~, \cr
\widetilde{\dpol2} =&~{25\over16}(3a_1 + 2a_2)\big(
           -{3\over2\sqrt 5}\del^2\phi -{1\over2}(\del\phi)^2 \big)
	~-~{25\over8}(3a_1 + 2a_2)\big(
           {3\over2\sqrt 5}i\del^2 X -{1\over2}(\del X)^2
               \big) \cr
		&~+{5\over96}(5a_1 + 6a_2)\big( -{\sqrt 5}\del^2\phi~+~
		    5(\del\phi)^2 \big)~, \cr
\dpol3 =&~-{5\over16}(3a_1 + 2a_2) \big({1\over\sqrt 5}\del^3\phi
             ~-~\del^2\phi\del\phi \big)
         +{5\over16}(3a_1 + 2a_2) \big({3\over2\sqrt 5}i\del^3 X
             ~-~\del^2 X \del X \big) \cr
         &~+{5\over24}a_1 \big( -{\sqrt 5\over2}\del^3\phi~+~
		   {3\over2}\del^2\phi\del\phi~+~{\sqrt 5\over2}(\del\phi)^3
		   \big)~+~{5\over48}(7a_1 + 6a_2)\del\phi
		       \big( {3\over2}i\del^2 X
             ~-~{\sqrt 5\over2}(\del X)^2 \big)~, \cr
\dpol4 =&~-{1\over8}(25a_1 + 14a_2)\bigl( {11\over12\sqrt 5}\del^4\phi
            -{1\over2}\del^3\phi\del\phi-{1\over2}(\del^2\phi)^2 \bigr)\cr
    	&-{1\over8}(55a_1 + 34a_2)\big( {3\over4\sqrt 5}i\del^4 X
            -{1\over2}\del^3 X\del X-{1\over2}(\del^2 X)^2	\big) \cr
	&~+{5\over48}(7a_1 + 6a_2)\big( -{\sqrt 5\over6}\del^4\phi
	   -\del^3\phi\del\phi + {\sqrt 5}\del^2\phi(\del\phi)^2
	   \big)~+~{5\over24}a_1{\sqrt 5}\del\phi\big(
	    {3\over2\sqrt 5}i\del^3 X - \del^2 X\del X \big) \cr
    &-{5\over32}(3a_1 + 2a_2)\big( -{4\sqrt 5\over3}\del^4\phi
	   + 6(\del^2\phi)^2 + 2{\sqrt 5} \del^2\phi(\del\phi)^2 \big) \cr
	   & -{5\over8}(3a_1 + 2a_2)\del^2\phi\big( {3\over2}i\del^2 X
	   - {\sqrt 5\over2} (\del X)^2 \big) \cr
	 &+{25\over8}(3a_1 + 2a_2)\big( {1\over2\sqrt 5}i\del^4 X
	     -{1\over2}\del^3 X \del X + {9\over20}(\del^2 X)^2
		 - {3\over2\sqrt 5}i\del^2 X(\del X)^2 + {1\over4}(\del X)^4
		  \big)~.   \cr}
$$

The expressions for $G_{-n}$ and ${\tilde G}_{-n}$ of eq. (47) for the current
$j^{(-2)}$ of the (2,7) model are the following:
$$
\eqalign{
G_{-2}=&(-\eta+{23\over 7}\epsilon)
        \big(-{5\over2\sqrt 7}\del^2\phi -{1\over2}(\del\phi)^2  \big)
        -{18\over 7}\epsilon
         \big({5\over2\sqrt 7}i\del^2 X -{1\over2}(\del X)^2\big) \cr
{\tilde G}_{-2}=&({2\over3}\eta-{10\over 7}\epsilon)
        \big(-{5\over2\sqrt 7}\del^2\phi -{1\over2}(\del\phi)^2  \big)
        +(\eta+{3\over 7}\epsilon)
         \big({5\over2\sqrt 7}i\del^2 X -{1\over2}(\del X)^2\big) \cr
G_{-3}=&(-\eta+{3\over7}\epsilon)\big({5\over2\sqrt 7}i\del^3 X
             ~-~\del^2 X \del X \big) \cr
G_{-4}=&\eta\bigl( {13\over12\sqrt 7}\del^4\phi
            -{1\over2}\del^3\phi\del\phi-{1\over2}(\del^2\phi)^2 \bigr)\cr
       &+(\eta-{24\over7}\epsilon)\big( {5\over4\sqrt 7}i\del^4 X
            -{1\over2}\del^3 X\del X-{1\over2}(\del^2 X)^2	\big) \cr
       &+\epsilon\bigl[ -{5\over6\sqrt 7}\del^4\phi-{1\over2}
         \del^3\phi\del\phi+\big(-{5\over2\sqrt 7}\del^2\phi
         -{1\over2}(\del\phi)^2\big)^2\bigr]  \cr
       &+3\epsilon\bigl[{5\over6\sqrt 7}i\del^4 X
         -{1\over2}\del^3 X\del X+\big({5\over2\sqrt 7}i\del^2 X
         -{1\over2}(\del X)^2\big)^2\bigr]   \cr
       &-2\epsilon\big(-{5\over2\sqrt 7}\del^2\phi -{1\over2}(\del\phi)^2
           \big)\big({5\over2\sqrt 7}i\del^2 X -{1\over2}(\del X)^2\big) \cr}
$$

\refout

 \end